\documentclass[pre,superscriptaddress,floatfix,twocolumn,showpacs]{revtex4}
\usepackage[dvips]{graphicx}
\begin{document}

\title{Wang-Landau molecular dynamics technique to search for low-energy
conformational space of proteins}
\author{Takehiro Nagasima}
\affiliation{%
Center for Information Biology and DNA Data Bank of Japan,\\
National Institute of Genetics,\\
Mishima 411-8540, Japan
}%
\author{Akira R. Kinjo}
\email{akinjo@protein.osaka-u.ac.jp}
\affiliation{%
Center for Information Biology and DNA Data Bank of Japan,\\
National Institute of Genetics,\\
Mishima 411-8540, Japan
}%
\affiliation{%
Department of Genetics, The Graduate University for Advanced Studies (SOKENDAI), Mishima 411-8540, Japan
}%
\affiliation{%
Institute for Protein Research, Osaka University, Suita, 565-0871, Japan
}%
\author{Takashi Mitsui}
\affiliation{%
Fujitsu Limited, 1-9-3, Nakase, Mihama, Chiba 261-8588, Japan
}%
\author{Ken Nishikawa}
\email{knishika@genes.nig.ac.jp}
\affiliation{%
Center for Information Biology and DNA Data Bank of Japan,\\
National Institute of Genetics,\\
Mishima 411-8540, Japan
}%
\affiliation{%
Department of Genetics, The Graduate University for Advanced Studies (SOKENDAI), Mishima 411-8540, Japan
}%
\date{\today}
\begin{abstract}
Multicanonical molecular dynamics (MD) is a powerful technique for
sampling conformations on rugged potential surfaces such as protein.
However, it is notoriously difficult to estimate the multicanonical
temperature effectively.
Wang and Landau developed a convenient method for estimating the density
of states based on a multicanonical Monte Carlo method.
In their method, the density of states is calculated autonomously during
a simulation.
In this paper, we develop a set of techniques to effectively apply the
Wang-Landau method to MD simulations.
In the multicanonical MD, the estimation of the derivative of the
density of states is critical.
In order to estimate it accurately, we devise two original improvements.
First, the correction for the density of states is made smooth by using
the Gaussian distribution obtained by a short canonical simulation.
Second, an approximation is applied to the derivative, which is based on
the Gaussian distribution and the multiple weighted histogram technique.
A test of this method was performed with small polypeptides, Met-enkephalin 
and Trp-cage, and it is demonstrated that Wang-Landau MD is consistent with 
replica exchange MD but can sample much larger conformational space.
\end{abstract}

\maketitle
\section{Introduction}
Computer simulation has been established as a technique for studying the 
systems with many degrees of freedom such as spin glasses and proteins.
The difficulties in studying such systems stem from the fact that the 
conformational 
space is very large and there exist a huge number of local minimum energy 
states.
For studying interesting phenomena or calculating quantities such as the 
transition of states or the free energy of the system, a global sampling 
of the conformational space and uniform sampling on energy space is desired.
Canonical simulations at high temperatures realize the global sampling but low 
energy states are poorly sampled.
At low temperatures, simulations can sample low energy states but the global 
search is difficult.
In order to overcome these difficulties, many generalized-ensemble algorithms 
have been developed such as the multicanonical algorithm~\cite{Berg-Neuhaus} 
and the replica exchange method (REM)~\cite{Hukushima-Nemoto,Sugita-Okamoto}.

The multicanonical method is based on artificial weight factor by which a 
simulation realizes a flat energy histogram.
This weight factor is not known \emph{a priori} and has to be determined by 
iterating simulations and estimations.
If the value of the weight factor is partially too large or small, 
then it causes to disturb a random walk in energy space.
Therefore, the multicanonical weight factor needs to be fairly accurate.
However, obtaining an accurate weight factor is often difficult and requires
a great deal of expertise.
In REM, a number of non-interacting replicas of the original system at 
different temperatures are simulated independently and simultaneously.
Every few steps the temperatures of pairs of replicas are exchanged, subject 
to a detailed balance condition.
As the system size increases, however, the required number of replicas also 
greatly increases.
In such a huge simulation system, it takes an unrealistically long CPU time 
for a replica to traverse the entire temperature range.

Wang and Landau~\cite{Wang-Landau} proposed a simulation method to sample 
conformational space efficiently.
In their method (hereafter referred to as the WL method), 
the weight factor is automatically estimated while the multicanonical 
simulation is performed. This method is originally based on a Monte Carlo (MC) algorithm.

In this paper, we focus on efficient sampling over the conformational space of 
an all-atom model of protein.
Molecular dynamics (MD) and MC are the standard techniques for all atom 
protein simulations.
It is recognized that the sampling efficiency of the MC method is comparable, 
or sometimes superior to MD for liquid simulations.
In protein simulations, however, MD shows about 1.5 times better sampling 
efficiency~\cite{Yamashita-etal}.
This difference is attributed to the inertia force term in MD, which does 
not exist in MC.

Recently, MD with the WL method was applied to simulations of 
solutions~\cite{Aberg-etal} in which the original WL algorithm was applied 
to MD simulations in a straightforward manner.
As we mention in detail later, the simple application of the WL method to MD 
does not work well for a protein system for a number of reasons.
First, in the multicanonical MD or related methods such as the WL MD,
not only the density of states but its derivative are required.
In the WL method, the density of states is rugged, so that its derivative is  
of poor accuracy. 
Second, as the WL method realizes quick random walk in energy space then for 
a polymer 
such as a protein, temperature changes occur in a time scale shorter than is 
necessary for a change of the structure of the whole molecule.
Eventually it takes a long time for a meaningful structural change to occur.
In this paper, we present techniques for implementing an efficient WL MD 
method to circumvent these problems.

This paper is organized as follows. We describe a detailed formulation of the 
Wang-Landau molecular dynamics (WLMD) method in Sec. \ref{sec:method}. 
In Sec. \ref{sec:model}, we introduce the model system for the present study, 
the Met-enkephalin, a 5-residue peptide, 
and the Trp-cage, a 20-residue protein. Next, we compare the sampling 
efficiency of the current method with that of the replica exchange MD (REMD) method 
in Sec. \ref{sec:result}.

\section{Method}
\label{sec:method}
\subsection{The Wang-Landau method}
Wang and Landau have developed a very powerful MC simulation 
technique for efficiently sampling conformational space.
In their method, the transition probability from energy $E_1$ to $E_2$ is 
given by
\begin{equation}
p(E_1\to E_2)=\min\left(\frac{n(E_1)}{n(E_2)}, 1\right),
\end{equation}
where $n(E)$ is the density of states.
This is the same as in the standard multicanonical MC method.
Since $n(E)$ is not known \textit{a priori}, one has to determine it by some 
means.
In conventional multicanonical simulations, it is determined by iterating 
short preliminary simulations using
\begin{equation}
\ln n^{(i+1)}(E)=\ln n^{(i)}(E)+\ln H^{(i)}(E) \qquad \forall E \in \mathcal{E},
\end{equation}
where $n^{(i)}(E)$ and $H^{(i)}(E)$ are the density of states and the histogram in the $i$th simulation, respectively, and $\mathcal{E}$ is a set of allowed 
energy levels.
What is unique of the WL method resides in the scheme 
for updating $n(E)$.
When an energy level, $E_i$, is visited at the time step, $i$, the existing 
$n(E)$ is modified by a modification factor $\gamma>1$, i. e.,
\begin{equation}
n^{(i+1)}(E)=\left\{
\begin{array}{ll}
\gamma n^{(i)}(E), & \mbox{if $E=E_i$}, \\
n^{(i)}(E), & \mbox{otherwise}.
\end{array}
\right.
\label{Eq:Wang-Landau}
\end{equation}
For a given energy, $E$, if $n^{(i)}(E)$ is smaller than the true density of 
state, $n(E)$, the energy state $E$ is sampled intensively so 
that $n^{(i)}(E)$ is updated to approach the true value, $n(E)$.
On the contrary, if $n^{(i)}(E)>n(E)$, the sampling of the energy state, 
$E$, is suppressed and other energy states are extensively sampled.
Although $n(E)$ is unknown at the very beginning of simulation, $n^{(i)}(E)$ 
approaches the true value quickly and automatically.
Since $n^{(i)}(E)$ is updated every step, it is possible for the system to 
escape local minima in a very short period of time.
\subsection{The Wang-Landau molecular dynamics}
We apply the WL method to MD simulations.
The equation of motion for the multicanonical MD is given by~\cite{Hansmann-etal,Nakajima-etal}
\begin{eqnarray}
m_k \frac{d \vec{v}_k}{d t} & = & -\frac{\beta_\mathrm{mu}}{\beta_0} \frac{\partial E}{\partial \vec{x}_k},\\
\label{Eq:motion1}
\beta_\mathrm{mu} & = & \frac{d \ln n(E)}{d E},
\label{Eq:betamu}
\end{eqnarray}
where $m_k$, $\vec{v}_k$, and $\vec{x}_k$ represent mass, velocity, and 
coordinate of $k$th atom, and the inverse temperature, 
$\beta_0=1/k_\mathrm{B}T_0$ ($k_\mathrm{B}$ is the Boltzmann constant), with 
the simulation temperature, $T_0$.
$\beta_\mathrm{mu}$ is referred to as the multicanonical temperature.
In a straightforward application of the WL method to MD, the density of states, 
$n(E)$, could be estimated according to Eq. (\ref{Eq:Wang-Landau}).
However, such a simple method does not work well in practice. 
It is necessary to accurately estimate not only $n(E)$ but also its 
derivative $\beta_\mathrm{mu}$.
In numerical calculation, $n(E)$ is divided into bins, and each bin is given 
a discrete value.
In general, the estimate of $\beta_\mathrm{mu}$ is of low accuracy due to 
the ruggedness between neighboring bins.
In order to smooth this ruggedness, we approximate canonical energy distribution at various temperatures by the Gaussian distribution and combine them with the weighted histogram method (WHAM)~\cite{Kumar-etal} for updating $n^{(i)}(E)$ and estimating the multicanonical temperature.

In the canonical ensemble at an inverse temperature, $\beta$, the distribution of energy $E$ is written as
\begin{equation}
  P_{\beta}(E) = n(E)e^{f-\beta E}, \label{Eq:canodis}
\end{equation}
where $n(E)$ is the density of states and 
\begin{equation}
  e^{-f} = \sum_En(E)e^{-\beta E}.\label{Eq:SINGLE2}
\end{equation}
The last equation [Eq. (\ref{Eq:SINGLE2})] defines the partition function.
The canonical distribution usually has a bell-like shape. At the energy $E_{\mathrm{peak}}$ corresponding to the peak of the distribution, the equation $d \ln P_\beta(E)/d E = 0$ is satisfied. This is rewritten as
\begin{equation}
\left. \frac{d \ln n(E)}{d E}\right|_{E=E_\mathrm{peak}}-\beta=0.
\label{Eq:canopeak}
\end{equation}
By comparing Eqs. (\ref{Eq:betamu}) and (\ref{Eq:canopeak}), one can see that 
the multicanonical temperature $\beta_\mathrm{mu}$ at energy $E$ is the
temperature of a canonical ensemble whose peak energy is $E$.
If $E_{\mathrm{peak}}$ and $n(E)$ are known, it is possible to solve 
Eq. (\ref{Eq:canopeak}) for $\beta$.
However, it is difficult to evaluate $E_\mathrm{peak}$ since $d \ln n(E)/d E$ is inaccurate if $n(E)$ is estimated from a simulation 
(which is the case in practice).
On the other hand, it is relatively easy to estimate the thermal average 
${\langle E\rangle}_\beta$ as
\begin{equation}
{\langle E\rangle}_\beta=\frac{\sum_E E n^{(i)}(E) e^{-\beta E}}{\sum_E n^{(i)}(E) e^{-\beta E}}, \label {eq:average}
\end{equation}
where $n^{(i)}(E)$ is the density of states estimated from the simulation.
The variance of the energy can be estimated in the same manner,
\begin{equation}
\sigma_\beta^2=\frac{\sum_E (E - \langle E \rangle_{\beta})^2 n^{(i)}(E) e^{-\beta E}}{\sum_E n^{(i)}(E) e^{-\beta E}}. \label{eq:variance}
\end{equation}
Hansmann~\cite{Hansmann} suggested using ${\langle E\rangle}_\beta$ instead 
of $E_\mathrm{peak}$. In his prescription, $\beta_\mathrm{mu}$ is determined 
from the table of ${\langle E\rangle}_\beta$.
The canonical distribution of the systems that we are interested in is 
similar to the Gaussian distribution.
Using ${\langle E\rangle}_\beta$ and the deviation $\sigma^2_\beta$ at $\beta$, 
we assume that the following approximation holds:
\begin{equation}
P_\beta(E)=\frac{\exp\left(\frac{-(E-{\langle E\rangle}_\beta)^2}{2\sigma^2_\beta}\right)}{\sum_{E'}\exp\left(\frac{-(E'-{\langle E\rangle}_\beta)^2}{2\sigma^2_\beta}\right)},
\label{Eq:gaussdis}
\end{equation}
which can be rearranged, using Eqs. (\ref{Eq:canodis}) and (\ref{Eq:gaussdis}),
to
\begin{equation}
n(E)=\frac{\exp\left(-\frac{(E-{\langle E\rangle}_\beta)^2}{2\sigma^2_\beta}+\beta E-f\right)}{\sum_{E'} \exp\left(-\frac{(E'-{\langle E\rangle}_\beta)^2}{2\sigma^2_\beta}\right)}.\label{Eq:single}
\end{equation}
It is possible to calculate the right-hand side from a canonical simulation, 
but the accuracy decrease as $\left( E-{\langle E\rangle}_\beta \right)^2$ 
increases.
In order to keep the accuracy for a wide energy range, $n(E)$ is estimated by using the WHAM technique with a number of temperatures,
\begin{equation}
n(E)=\frac{\sum_j w_j P_{\beta_j}(E)}{\sum_j w_j e^{f_j-\beta_j E}},
\label{Eq:WHAM1}
\end{equation}
where
\begin{equation}
  \label{Eq:WHAM2}
 e^{-f_j} = \sum_E n(E)e^{-\beta_j E}  
\end{equation}
and $w_j$'s are appropriate weight factors (see below). $n(E)$ and $f_j$ 
can be obtained by solving Eqs. (\ref{Eq:WHAM1}) and (\ref{Eq:WHAM2}) iteratively.
From Eqs. (\ref{Eq:gaussdis}) and (\ref{Eq:WHAM1}), it follows that
\begin{equation}
\frac{d \ln n(E)}{d E}=\frac{\sum_j w_j\left( \beta_j n(E) e^{f_j-\beta_j E}-P_{\beta_j}(E) \frac{E-{\langle E\rangle}_{\beta_j}}{\sigma_{\beta_j}^2}\right)}{\sum_j w_j P_{\beta_j}(E)},
\label{Eq:dlogn1}
\end{equation}
which can be further reduced, by using Eq. (\ref{Eq:canodis}), to
\begin{equation}
\frac{d \ln n(E)}{d E}=\frac{\sum_j w_j P_{\beta_j}(E) \left(\beta_j-\frac{E-{\langle E\rangle}_{\beta_j}}{\sigma_{\beta_j}^2}\right)}{\sum_j w_j P_{\beta_j}(E)}\equiv\beta (E).\label{Eq:dlogn2}
\end{equation}
Both $n(E)$ and $f_j$ vanish in this expression. We use this equation for estimating the multicanonical temperatures.
Note that Eq. (\ref{Eq:gaussdis}) is not exact.
This implies that our method may not be suitable for a system with doubly
 peaked distribution, for example. Nevertheless, such an error should be small if we use a sufficiently large number of temperatures.

In a WLMD simulation, we need to specify the lowest and highest temperatures for the system ($\beta_{\min}$ and $\beta_{\max}$, respectively), 
and prepare energy bins and temperatures corresponding to each energy bin. 
The initial estimate of the density of states, $n^{(1)}(E)$, is obtained as 
follows.
We first run a short (say, 100 steps) canonical MD simulation at the inverse temperature $\beta_{\max}$ (corresponding to the highest temperature),
and calculate the average and variance of the energy from this short simulation.
We then apply Eqs. (\ref{Eq:single}) and (\ref{Eq:SINGLE2}) iteratively, which 
yields the initial estimate $n^{(1)}(E)$.
The rest of a WLMD proceeds as follows.
\vspace{1ex}\\

(1) Set $i = 1$ and $\beta_\mathrm{mu}^{(i)} = \beta_{\max}$.

(2) A canonical MD is performed for a very short period of time (say, 100 steps) according to Eq. (\ref{Eq:motion1}) with fixed $\beta_{\mathrm{mu}}^{(i)}$.
During this period, the time average $\bar{E}_{\beta_{\mathrm{mu}}^{(i)}}$ and the variance $\sigma_{\beta_{\mathrm{mu}}^{(i)}}^2$ of 
the energy are estimated. The Gaussian energy distribution based on these 
values is denoted $P_{\beta_{\mathrm{mu}}}^{(i)}(E)$. Let the energy value of the last step of this canonical simulation be $E_{\mathrm{last}}^{(i)}$.
Let $E_{\mathrm{mu}}^{\min,i} = \min_{k=1,\dots,i}\bar{E}_{\beta_{\mathrm{mu}}^{(k)}}$
and $E_{\mathrm{mu}}^{\max,i} = \max_{k=1,\dots,i}\bar{E}_{\beta_{\mathrm{mu}}^{(k)}}$.

(3) The average and variance of energy at the inverse temperature 
$\beta_{\max}$ are estimated using all the energy values that have been 
visited with $\beta_{\mathrm{mu}} = \beta_{\max}$ up to this time step. Using these values, the 
Gaussian distribution $P_{\beta_{\max}}^{(i)}(E)$ is obtained. Similarly, the 
Gaussian distribution at $\beta_{\min}$, 
$P_{\beta_{\min}}^{(i)}(E)$, is estimated. $P_{\beta_{\min}}^{(i)}(E)$ is set 
to zero if the simulation has never visited this temperature.

(4) For each energy bin, $E_j$, calculate the corresponding temperature 
$\beta_{j}^{(i)} = \beta^{(i)}(E_j)$ according to Eq. (\ref{Eq:dlogn2}) with the Gaussians 
$P_{\beta_j^{(i-1)}}(E)$, $P_{\beta_{\max}}^{(i)}(E)$, $P_{\beta_{\min}}^{(i)}(E)$,
and $P_{\beta_{\mathrm{mu}}}^{(i)}(E)$. The weight $w_j$ is set to 1 except for 
$P_{\beta_{\mathrm{mu}}}^{(i)}(E)$ for which the weight is set to a specified 
modification factor $w_{\mathrm{mu}} > 0$ [this factor corresponds to $\gamma$ in
Eq. (\ref{Eq:Wang-Landau})].
(Note that, for $i=1$, $P_{\beta_j^{(i-1)}}(E)$ is not yet defined and is set to zero.)

(5) Using the temperatures $\beta_{j}^{(i)}$ obtained in the 
previous step, calculate, from Eqs. (\ref{eq:average}) and 
(\ref{eq:variance}) as well as the current estimate of the density of states $n^{(i)}(E)$, 
the average and variance, $\langle E \rangle_{\beta_{j}^{(i)}}$ and 
$\sigma_{\beta_{j}^{(i)}}^{2}$, and the Gaussian distribution, $P_{\beta_{j}^{(i)}}(E)$.

(6) Iterate steps 4 and 5 until all the $\beta_{j}^{(i)}$'s converge.

(7) Update the density of states using Eqs. (\ref{Eq:WHAM1}) and (\ref{Eq:WHAM2})
and the Gaussians
to obtain the next estimate $n^{(i+1)}(E)$.

(8) Estimate the next multicanonical temperature $\beta_{\mathrm{mu}}^{(i+1)}$ 
according to Eq. (\ref{Eq:dlogn2}) at the energy of the last canonical MD step,
$E_{\mathrm{last}}^{(i)}$.
If $E_{\mathrm{last}}^{(i)} < E_{\mathrm{mu}}^{\min,i}$ 
or $E_{\mathrm{last}}^{(i)} > E_{\mathrm{mu}}^{\max,i}$,         
$\beta_{\mathrm{mu}}^{(i+1)}$ is evaluated at $E_{\mathrm{mu}}^{\min,i}$ or 
$E_{\mathrm{mu}}^{\max,i}$, respectively. This is necessary for a stable simulation.

(9) Set $i=i+1$ and go back to step 2.
\vspace{1ex}\\

In order to avoid numerical overflow in the calculation of $n^{(i+1)}(E)$ in 
the step 7, we limit the summation in Eq. (\ref{Eq:WHAM1}) to those inverse 
temperatures $\beta_{j}^{(i)}$ satisfying 
\begin{equation}
  \label{eq:limitsum}
 E_{\mathrm{mu}}^{\min,i}\leq \langle E \rangle_{\beta_{j}^{(i)}} \leq   
E_{\mathrm{mu}}^{\max,i}.
\end{equation}
Recently, Kim \emph{et al. }~\cite{Kim-etal} developed the method to 
determine the multicanonical temperature automatically.
The notable points in their method are the short period update and adding the 
derivative of histogram to the multicanonical temperature.
While their method is different from our method in the way to update the 
histogram, the idea of the short term update is common.
They use the finite difference of raw data (energy histogram) for update.
Since the present method uses the Gaussian mask and the WHAM technique to 
estimate the density of states and its derivative, it is expected to be more
robust. 
\subsection{Separation of bond and non-bond interaction}
When we first applied the above method to a protein system, 
a random walk in the energy space was readily realized.
However, the protein conformation changed very little in the whole process.
It was found that the energy change predominantly originated from the bond 
length and the bond angle deviations.
As a result, the bond energy almost solely contributed to $n(E)$.
To avoid this artifact, we separated the bond interaction from the non-bond 
interaction, and applied the multicanonical ensemble only to the non-bond 
interaction.
The energies of bond and non-bond interactions are written as 
$E_\mathrm{b}$ and $E_\mathrm{n}$, respectively, i. e.,
\begin{equation}
E=E_\mathrm{b}+E_\mathrm{n}.
\end{equation}
Using $E_\mathrm{b}$ and $E_\mathrm{n}$, density of state is represented 
as $n(E_\mathrm{b}, E_\mathrm{n})$.
The equations of motion with the $n(E_\mathrm{b}, E_\mathrm{n})$ is now 
expressed as
\begin{equation}
m_k \frac{d \vec{v_k}}{d t}=-\frac{\beta_\mathrm{b}}{\beta} \frac{\partial E_\mathrm{b}}{\partial \vec{x}_k}-\frac{\beta_\mathrm{n}}{\beta} \frac{\partial E_\mathrm{n}}{\partial \vec{x}_k},
\end{equation}
with
\begin{eqnarray}
\beta_\mathrm{b}&=&\frac{\partial \ln n(E_\mathrm{b}, E_\mathrm{n})}{\partial E_\mathrm{b}},\\
\beta_\mathrm{n}&=&\frac{\partial \ln n(E_\mathrm{b}, E_\mathrm{n})}{\partial E_\mathrm{n}},
\end{eqnarray}
Then it is possible to give different temperatures for bond and non-bond 
interactions.
In order to suppress the motion due to the bond interaction and to facilitate
 global conformational changes, $\beta_\mathrm{b}$ is fixed 
whereas $\beta_\mathrm{n}$ is varied according to the WLMD scheme from a low 
to high temperature.

\section{Model for Numerical Calculation}
\label{sec:model}

We carried out simulations using a customized version of the PRESTO molecular 
simulation package~\cite{Morikami-etal}, in which the all-atom version of the 
AMBER force-field parameters (C96)~\cite{Kollman-etal} was used.
\subsection{Met-enkephalin}
\label{sec:enke}
We used the Met-enkephalin [Protein Data Bank (PDB) code: 1PLW; amino acid sequence: YGGFM] as a model system for checking the correctness of the method.
Eight independent WLMD simulations were conducted for 10 ns (80 ns in total), 
starting from the extended conformation with the temperature ranging between 
200 and 700 K. No solvent effect was included but a distance-dependent dielectric constant was used ($\epsilon = 4r$, where $r$ is the inter-atomic distance).
The density of states was updated every 1000 steps (unit time step was 0.5 fs).
The objective of these simulations is to check whether the present method 
can yield a plausible density of states compared to the REMD simulations.
Therefore, the modification factor for the density of states was initially set 
to $w_{\beta_\mathrm{mu}} = 12.8$ which was decreased by the factor of 0.8 every 1 ns.
This procedure is analogous to that proposed originally by Wang and Landau \cite{Wang-Landau}.

For comparison, we carried out an extensive REMD simulation with eight replicas, 
each running 100 ns. Temperatures ranged between 200 and 700 K, and were 
distributed exponentially. Replica-exchanges were tried every 20 steps.
The average exchange acceptance ratio was approximately $18.6$\%.
For the calculation of principal components and density of states, those 
conformations generated during the first 1 ns of the simulation were discarded.
\subsection{Trp-cage}
\label{sec:trp-cage}

The protein used for demonstrating the efficiency of the method is a 20-residue 
protein, Trp-cage (PDB code: 1L2Y; amino acid sequence: 
NLYIQ WLKDG GPSSG RPPPG), which was derived from the C-terminal fragments of a 
39-residue exendin-4 peptide~\cite{Neidigh-etal}.
Physicochemical studies showed that Trp-cage folds spontaneously and 
cooperatively.
It contains a short $\alpha$-helix in residues 2-9, a $3_{10}$-helix in 
residues 11-14, and a C-terminal polyproline II helix to pack against the 
central tryptophan (Trp-6). Several simulation studies on this Trp-cage have 
already been published ~\cite{Simmerling-etal,Chowdhury-etal,Zhou,Ota-etal}.

An implicit solvent model~\cite{Ooi-etal} was used with the dielectric 
constant was set equal to $4r$, where $r$ is distance between point 
charges.
Furthermore, we employed an empirical dihedral angle potential which was 
developed in our laboratory~\cite{Kinjo-etal}.
The simulations were started from an extended conformation.
The unit time step was set to 0.5 fs, and we ran MD simulations for 10 ns.
We set $T_\mathrm{min} = 1/k_{B}\beta_{\min} =200$ K, $T_\mathrm{max} = 1/k_B\beta_{\max} = 700$ K, $T_\mathrm{b}=300$ K, 
the multicanonical temperature was updated every 400 steps (0.2 ps).
The number of energy bins was set to 256.
We carried out 24 runs with different random initial velocities.
The modification factor for the density of states is set such that 
$w_{\beta_{\mathrm{mu}}} = 12.8$.

For comparison, an alternative simulation with 
REMD was performed under the equivalent condition, that is, 
24 replicas each running for 10 ns. 
Temperatures exponentially varied from 200 to 700 K 
and swapping was carried out every 400 steps.
The average acceptance ratio was $33.1$\%.

The run time required for one WLMD simulation was 
184--193 h (189 h on average) on an Intel Pentium III (1GHz) processor. 
For REMD, the run time per CPU was 207 h.
WLMD seems slightly more efficient than REMD in spite of the extra effort 
required for the WHAM iterations. The lower computational efficiency of REMD
is caused by the synchronization process which is necessary for different 
replicas to exchange.

\section{Results and Discussion}
\begin{figure}[tb]
  \centering
  \includegraphics[width=8cm]{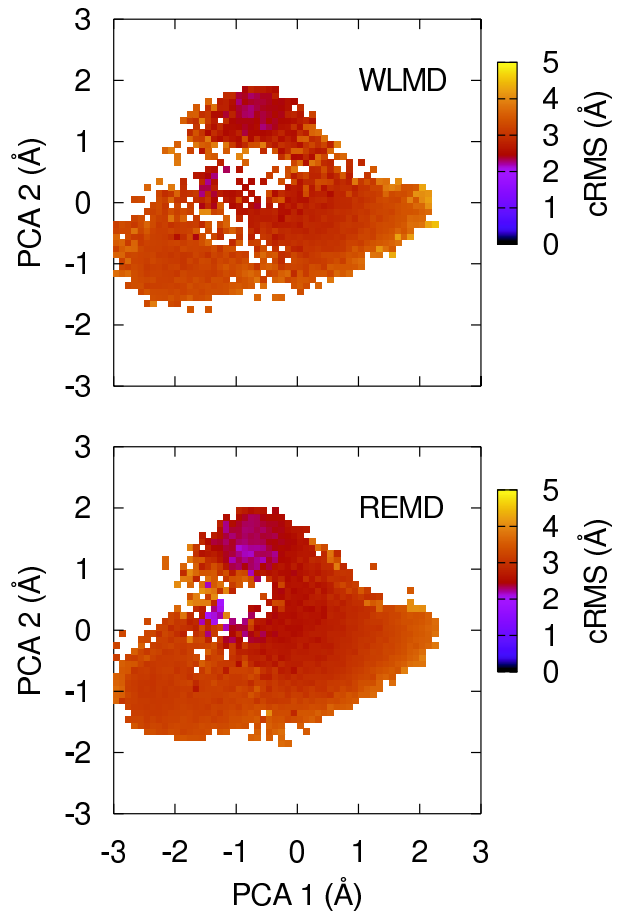}
  \caption{Conformations of Met-enkephalin sampled by Wang-Landau molecular 
dynamics (WLMD, upper panel) and replica-exchange molecular dynamics (REMD, lower panel) projected onto the first two principal component axes (PCA). During WLMD 160\,000 conformations were saved and 10\,000 lowest energy conformations were selected. In the same manner, 10\,000 lowest energy conformations were selected for REMD. These 20\,000 conformations in total were used for determining the principal axes. Upper panel: 10\,000 lowest energy conformations obtained by WLMD are projected. Lower panel: 100\,000 lowest energy conformations obtained by REMD are projected. Each pixel is colored according to the cRMS from the PDB structure.}
  \label{fig:enk-pca}
\end{figure}

In order to check the correctness of WLMD, we first compare the results of 
WLMD and REMD for a small peptide Met-enkephalin. Figure \ref{fig:enk-pca} shows
that the conformational spaces sampled by the two methods closely overlap, 
indicating that WLMD can sample the conformational space as widely as REMD.
However, it should be noted that 10 times more points are plotted for REMD than WLMD: if we plot the same number of points, the REMD result looks 
much more sparse.
\label{sec:result}

\begin{figure}[tb]
  \centering
  \includegraphics[width=8cm]{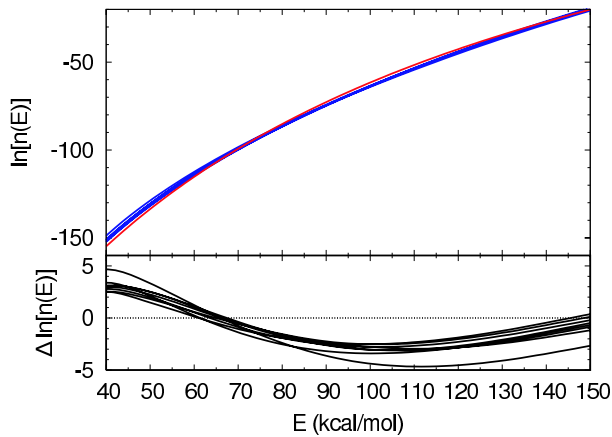}
  \caption{Comparison of density of states of Met-enkephalin obtained by WLMD (blue lines) and REMD (red line). Eight density of states were obtained from eight WLMD 
simulations (blue lines), one density of states was calculated from one REMD 
simulation with eight replicas (red line). The lower panel shows the difference 
of each density of states of WLMD from that of REMD.
}
  \label{fig:enk-dos}
\end{figure}

For a WLMD simulation to work at all, good estimate of 
the density of states should be obtained as the simulation proceeds. 
We compare the density of states obtained by eight WLMD simulations with that 
obtained by a long REMD simulation in Fig. \ref{fig:enk-dos}.
Again, we see that the results of the two methods agree quite well.
A close examination reveals that WLMD slightly overestimates the density 
of low energy ($<70$ kcal/mol) states 
(Fig. \ref{fig:enk-dos}, lower part) compared to REMD. 
Nevertheless, such difference is 
of a small fraction ($\approx 3\%$) of the entire range of the density of states. Since WLMD is simply a method to efficiently estimate the density of states 
for multicanonical MD, we can always further refine the density of states using
an ordinary multicanonical MD procedure.

We next turn to our main application of WLMD: sampling of low-energy 
conformations of Trp-cage.
\begin{figure}[tb]
  \centering
  \includegraphics[width=8cm]{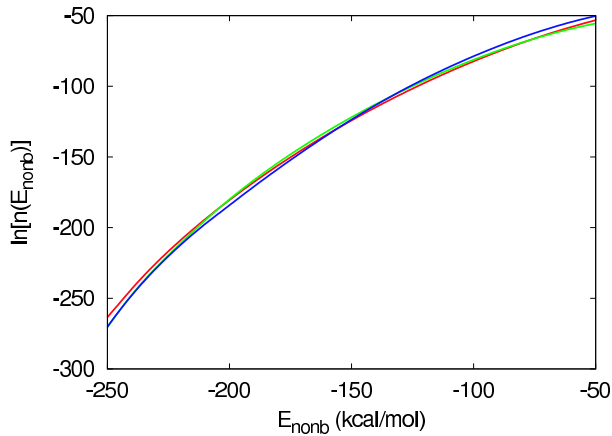}
  \caption{Density of states of Trp-cage obtained from three independent Wang-Landau molecular dynamics simulations. Each line represents the density of states obtained from one WLMD run.}
  \label{fig:dos}
\end{figure}
In order to check if the density of states is still accurately estimated for this bigger protein, we plotted the density of states obtained by three 
independent WLMD simulations (Fig. \ref{fig:dos}). Although we cannot know 
the exact density of states, the consistency among these estimates of 
density of states suggests 
that the present scheme is at least self-consistent. 
In order to calculate thermodynamic quantities,
one should gradually decrease the value of $w_{\mathrm{mu}}$ toward zero and 
set $w_{\mathrm{mu}} = 0$ for the production run, as 
suggested by Wang and Landau~\cite{Wang-Landau}.
In the present example with Trp-cage, our purpose is efficient sampling of 
low energy conformations but not calculating thermodynamic quantities so 
that $w_{\mathrm{mu}}$ is always kept greater than 0.
\begin{figure}[tb]
  \centering
  \includegraphics[width=8cm]{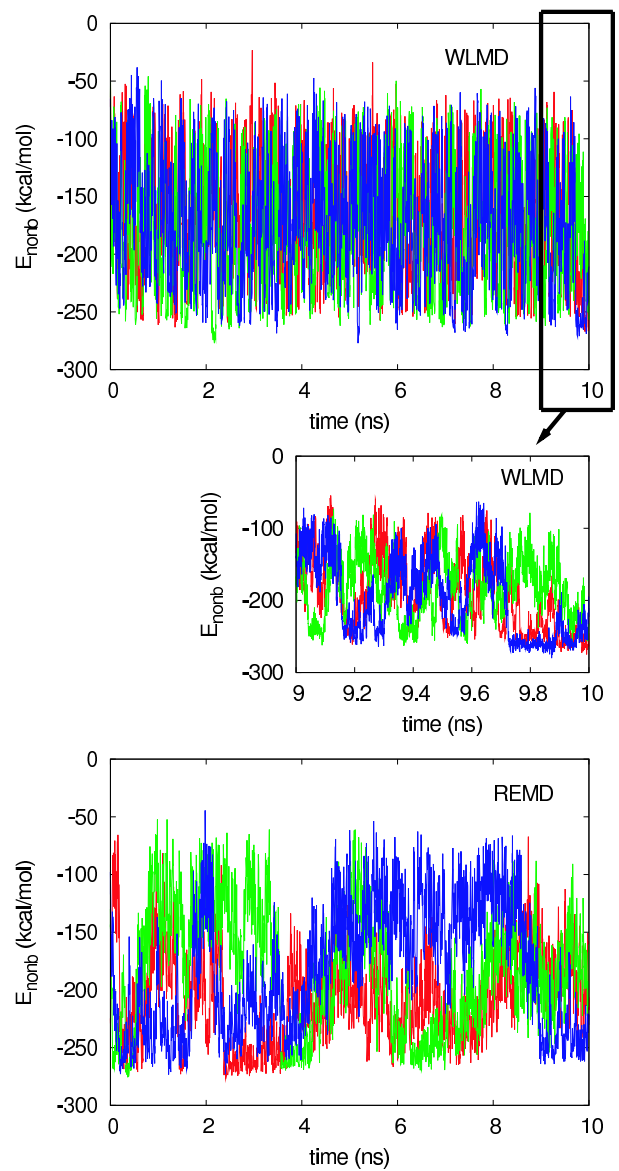}
  \caption{Time series of non-bond energy of Trp-cage. Upper panel: Wang-Landau molecular dynamics; the trajectories for the last 1 ns are enlarged for clarity. Lower panel: Replica exchange molecular dynamics. Arbitrarily selected three trajectories out of 24 are shown.}
  \label{fig:ene}
\end{figure}
We next compare the (non-bond) energy trajectories of WLMD and REMD simulations (Fig. \ref{fig:ene}). We selected three trajectories arbitrarily from 24 simulations or replicas. It is immediately apparent that the WLMD trajectories 
(Fig. \ref{fig:ene}, upper panel) traverse a wide range of energy very rapidly 
compared to
the REMD trajectories (Fig. \ref{fig:ene}, lower panel). In fact, one 
trajectory of REMD seems to be trapped in a local minimum.

\begin{figure}[tb]
  \centering
  \includegraphics[width=8cm]{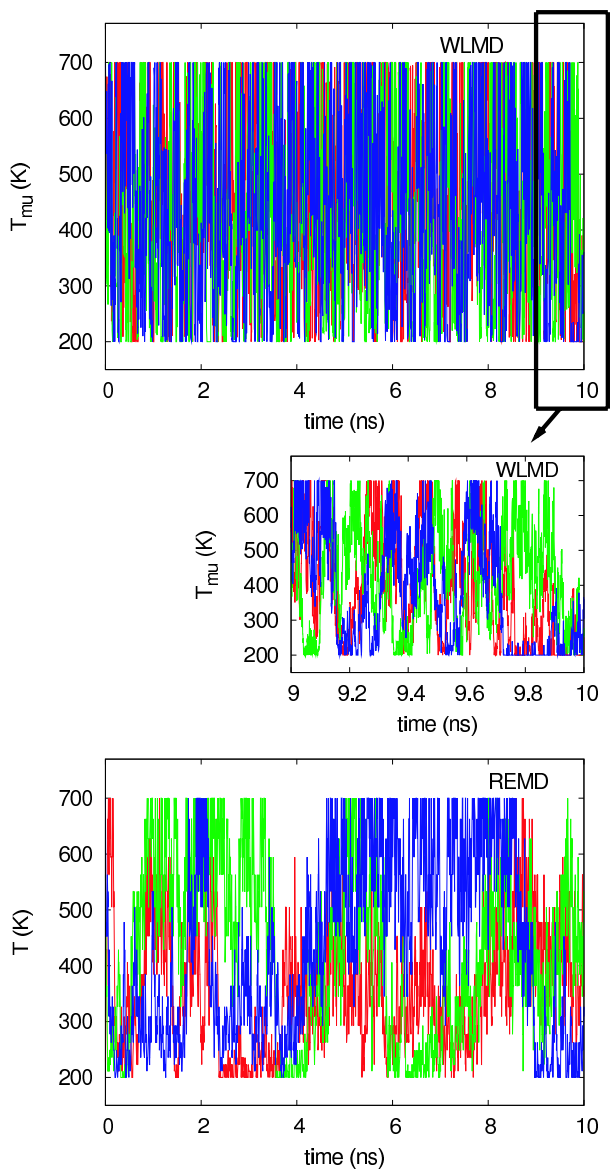}
  \caption{Time series of temperatures of Trp-cage. 
Upper panel: Three trajectories from Wang-Landau molecular dynamics 
simulations; the trajectories for the last 1 ns are enlarged for clarity. 
Lower panel: Trajectories of three replicas from replica exchange molecular dynamics simulations.
In both figures, each trajectory corresponds to that of Fig. \ref{fig:ene}.}
  \label{fig:temp}
\end{figure}
A similar trend is also observed for temperature trajectories 
(Fig. \ref{fig:temp}).
The multicanonical temperature changes very rapidly but smoothly in WLMD
(Fig. \ref{fig:temp}, upper panel) while the temperature changes rather slowly 
in REMD (Fig. \ref{fig:temp}, lower panel).
One reason for such a behavior in WLMD is that the multicanonical temperature 
can vary continuously between $T_{\min}$ (200 K) and 
$T_{\max}$ (700 K) as it is determined by Eq. (\ref{Eq:dlogn2}).
On the other hand, in REMD, there are only 24 allowed and fixed temperatures 
if we use 24 replicas so that the acceptance ratio of temperature swapping can 
be arbitrarily small. 

\begin{figure}[tb]
  \centering
  \includegraphics[width=8cm]{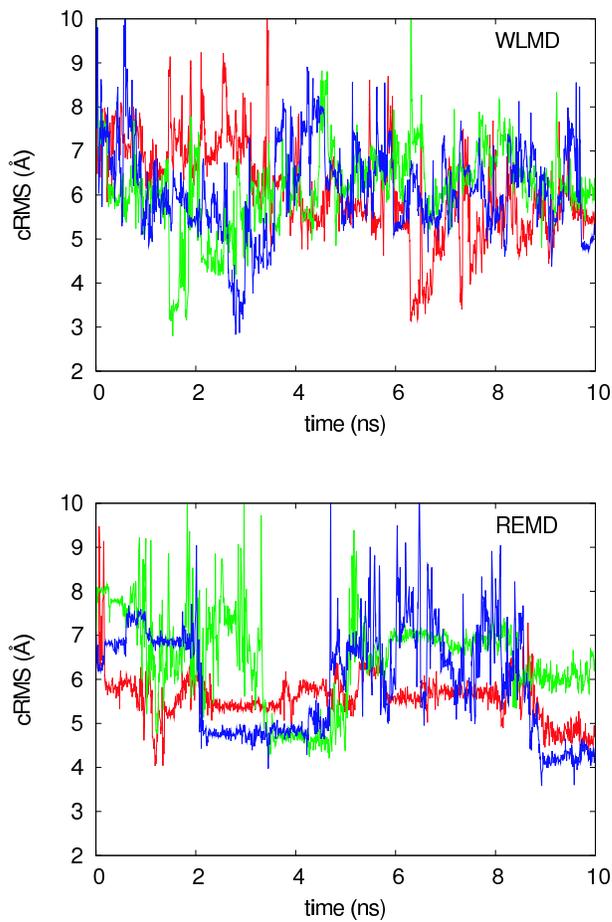}
  \caption{Trajectories of coordinate root mean square deviations (cRMS) from the native structure of Trp-cage.
Upper panel: WLMD; lower panel: REMD. 
Each trajectory corresponds to that of Fig. \ref{fig:ene}.}
  \label{fig:crms}
\end{figure}

Although rapid traversal of a wide energy or temperature range is necessary,
it is not sufficient for efficient sampling of conformational space.
In fact, rapid energy change may be associated with very localized motions such
as bond stretching. In order to examine the sampling efficiency of WLMD, we 
first checked the trajectories of the coordinate root mean square deviation 
(cRMS) from the native structure of Trp-cage (Fig. \ref{fig:crms}).
As expected, cRMS changes more slowly (Fig. \ref{fig:crms}) than might be suggested by the energy trajectories (Fig. \ref{fig:ene}) in both WLMD and REMD.
Nevertheless, WLMD is much less prone to being trapped in local minima and 
conformations appear to keep moving compared to REMD. A similar trend was observed for the trajectories of end-to-end distance (not shown).

\begin{figure}[tb]
  \centering
  \includegraphics[width=8cm]{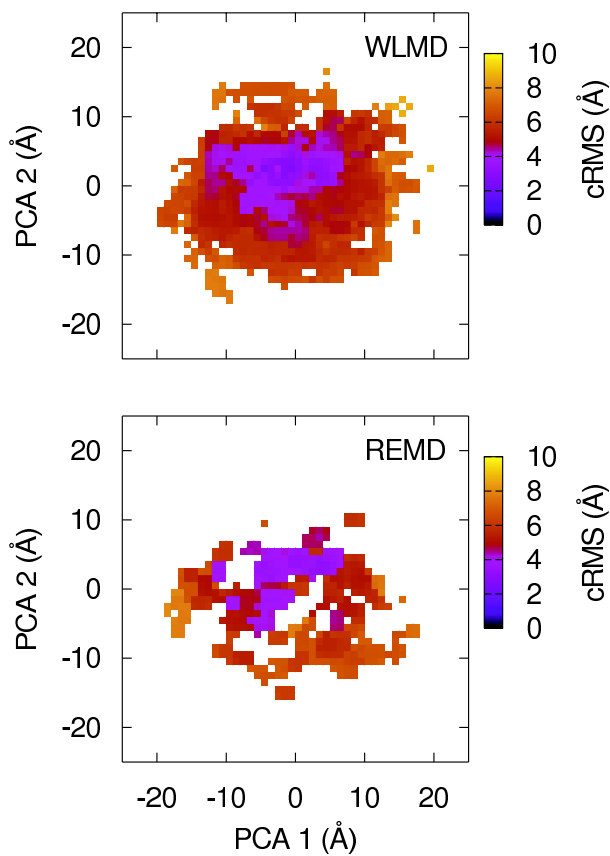} 
  \caption{Conformations of Trp-cage projected onto the principal space spanned by the first and second principal axes. Upper panel: WLMD; lower panel: REMD. Each pixel is colored according to the cRMS deviation from the native structure of Trp-cage.}
  \label{fig:pca}
\end{figure}

In order to visualize the efficiency of conformational sampling, we carried out
principal component analysis~\cite{Kitao-Go}. Out of 1\,200\,000 conformations saved during 24 WLMD
simulations, 50\,000 lowest energy conformations were extracted, to which
50\,000 lowest energy conformations out of 1\,200\,000 saved during the REMD 
simulation (24 replicas) were added. These 100\,000 conformations were used to 
define the principal axes. Conformations sampled in WLMD and REMD were then 
projected onto the principal space (Fig. \ref{fig:pca}).
It is apparent from Fig. \ref{fig:pca} that WLMD covers much more space than 
REMD. To confirm this observation more quantitatively, we divided the space
spanned by the first three principal axes (each from -25 to 25 \AA) into 
small cells (1.0 $\times$ 1.0 $\times$ 1.0 \AA$^3$), and counted the number 
of occupied cells. Out of 125\,000 cells, 4\,335 (including the cell containing 
the native structure) were occupied by WLMD while only 1\,174 by REMD and the 
native structure cell was not occupied. 
Therefore, in terms of the cell occupancy, WLMD is 3.7 times 
more efficient than REMD for sampling protein conformations.

In summary, we formulated the Wang-Landau molecular dynamics method.
In so doing, we adopted the Gaussian masking for updating the density of states,
and developed a technique to reliably estimate the multicanonical temperature.
It was shown that the WLMD method indeed samples conformational space of a 
protein more efficiently than the replica exchange MD method. 
Apart from the inaccuracy in the molecular force field, the present method 
will serve as a useful tool for simulation studies of protein molecules.

\begin{acknowledgments}
  This work was supported in part by a grant-in-aid from the MEXT, Japan.
\end{acknowledgments}
%
%

\end{document}